\documentclass[12pt]{article}
\usepackage{amssymb,amsmath,epsfig}

\numberwithin{equation}{section}

\begin{document}

\title{\bf Higher Dimensional Inhomogeneous Perfect Fluid Collapse in \emph{f(R)} Gravity}

\author{G. Abbas $^1$\thanks{abbasg91@yahoo.com;ghulamabbas@iub.edu.pk},M. S. Khan$^2$,
Zahid Ahmad$^2$\thanks{zahidahmad@ciit.net.pk}, M. Zubair$^3$ \thanks{mzubairkk@gmail.com} \\
$^1$Department of Mathematics, The Islamia\\
University of Bahawalpur, Bahawalpur, Pakistan.\\
$^2$ Department of Mathematics,\\
COMSATS Institute of Information Technology,\\
Abbottabad, KPK, Pakistan\\
$^3$ Department of Mathematics,\\
COMSATS Institute of Information Technology,\\
Lahore, Pakistan }
\date{}

\maketitle

\begin{abstract}
 This paper is about the $n+2$-dimensional gravitational contraction of inhomogeneous fluid without heat flux in the framework of $f(R)$ metric theory of gravity. Matching conditions for two regions of a star has been derived by using the Darmois junction conditions. For the analytic solution of equations of motion in modified $f(R)$ theory of gravity, we have taken scalar curvature as constant. Hence final result of gravitational collapse in this frame work is the existence of black hole and cosmological horizons, both of these form earlier than singularity. It has been shown that constant curvature term $f(R_{0})$ ($R_0$ is constant scalar curvature) slows down the collapsing process.

\end{abstract}
\section{\protect \LARGE Introduction}

Recently, the modified theories of gravity have attracted the attention of many researchers in theoretical and observational cosmology and astrophysics. One of most active research direction is the exploration of many astrophysical problems in modified \emph{f(R)} metric theory of gravity. It is most reliable to consider this theory due to its simplicity in derivation as it is obtained by taking Ricci scalar, $R$ action as a function \emph{f(R)} in the Einstein-Hilbert gravitational action. All the modifications of General Relativity (GR) explore the problem of dark energy in a more scientific way \cite{1}-\cite{3}. Some major modifications of GR are $f(R)$ gravity \cite{2}, $f(R,\mathcal{T})$ gravity ($\mathcal{T}$ is the
trace of energy-momentum tensor $\mathcal{T}_{\alpha\beta}$) \cite{4},
the $f(R,\mathcal{T},\mathcal{Q})$ gravity (where $\mathcal{Q}=R_{\alpha\beta}T^{\alpha\beta}$) \cite{5},
Gauss-Bonnet gravity \cite{5aa}, teleparallel modified theories \cite{6}, scalar-tensor theories  \cite{7}.
 It has been pointed out by many researcher \cite{3}-\cite{6} that $f(R)$ theory confirms that when major interactions are unified they leads to an actions, which involves curvature invariants of nonlinear order.

During the last decades many renowned researchers have investigated the gravitational collapse in modified theories.  It has been shown that one can goes beyond the general relativity, the more has chances of admitting an uncovered singularity. A lot of work has been done in GR about the gravitational collapse \cite{8zz}-\cite{20zz}. The non-linear electrodynamics static black holes (BH) solutions have been formulated within the frame work of $f(R)$ \cite{f111}. In this connection Borisov et al.\cite{f1} have investigated the spherical gravitational collapse in $f(R)$ gravity by performing one dimensional numerical simulations . In this study, the non-linear self-interaction coupling of scalar field has been included in the dynamical equations and a relation scheme has been used to follow the gravitationally contracting solutions. During the scalar field collapse in $f(R)$ gravity, the density increases rapidly near the viral radius, which may provide the observable test of gravity. Schmidt et al.\cite{f2} have used large scale simulations for spherical collapse in $f(R)$ gravity to estimated the halo mass function. Capozziello et al.\cite{f3} have investigated the hydrostatic equilibrium and stellar structure in $f(R)$ gravity.

Cembranos et al.\cite{f4}, have explored the gravitational collapse of matter with uniform density in $f(R)$ gravity. This analysis provide the information about  the structure formation in the early universe. It has been remarked that for some particular models of $f(R)$ gravity, the gravitational collapse process would help to constraint the models that exhibit the late-time cosmological acceleration. The time of collapse in this frame work has been noted much smaller as compared to the age of the universe while it is much longer to form the matter clustering.  All, the previous investigation of gravitational collapse in $f(R)$ gravity imply that gravity is highly attractive force this is in the agreement with the observed consequences of $f(R)$ gravity. It is an admitted fact that a scalar force reduce the time for gravitational collapse because of its attractive behavior. Ghosh and Maharaj \cite{f5} have explored the exact models of null dust collapse in metric $f(R)$ with the constant scalar curvature condition. Further, in this situation the null dust collapse leads to the formation of naked singularities, hence violating the cosmic censorship conjecture (CCC) in $f(R)$ gravity. Goswami et al.\cite{f6} has been proved that gravitational collapse of heat conducting, shearing and anisotropic fluid in $f(R)$ seems to be unstable with respect to matter perturbations, there exists no apparent horizons and hence there occurs naked singularities. It is important to note that investigating CCC in modified gravity may be more complicated than in GR. As it is well established that inhomogeneity is closely related to related to spacetime shear and Weyl curvature of the collapsing star, which produces the naked singularities.

 The Oppenheimer-Snyder-Datt model \cite{8zz,f7}, which is widely acceptable model for BH formation via dynamical collapse is no longer a viable model in modified $f(R)$ gravity. Hence in order to establish the existence of BH solutions via stellar collapse in modified $f(R)$ theories, we have to find some new physically reasonable solutions in modified $f(R)$ gravity that may predict BH. Hwang et al. \cite{f8} have investigated the collapse of a charged BH in $f(R)$ gravity using double null formulism and constant scalar curvature assumption. In such charged BH solutions there appears a new type of singularity due to higher curvature corrections, the so-called $f(R)$ induced singularity. Pun et al. \cite{f9} have confirmed the existence of Schwarzschild like BH solution in $f(R)$ gravity. The modified $f(R)$ theories of gravity provide toy models for the existence and stability of neutron stars \cite{f10}. The stars which satisfy the baro-tropic equation of state $(\rho=\omega p)$ in $f(R)$ gravity, when undergos to gravitational collapse, the end state of the collapse would be a naked singularity violating CCC \cite{f11}.

Sharif and Nasir \cite{f12} have discussed the stability of expansion-free axially-symmetric fluids in $f(R)$ gravity. The gravitating source preserve its axial symmetry due to $f(R)$ extra degree of freedom. Also, the axially symmetric solutions have been formulated in $f(R)$ gravity by using the Newman-Janis method \cite{f13}. The rotating black string solutions have been investigated in $f(R)$-Maxwell gravity \cite{f14} using the constant scalar curvature assumption. Sharif and his collaborators \cite{f15}-\cite{f181} have studied the dynamical stabilities of many gravitating stellar systems in $f(R)$ theories with general form of $f(R)$ models. The instability and anti-evaporation of Reissner-Nordstr$\ddot{o}$m BH have been explored in the modified $f(R)$ gravity \cite{f19}.
The inhomogeneous dust as well perfect fluid collapse with the geodesic flow condition in $4D$ have been explored in \cite{11,12}. It has been remarked that $f(R)$ with constant scalar curvature would appears as an alternative to cosmological constant.

Recent advancements in string theory and other field theories indicate that gravity is a higher-dimensional interaction. It would be  interesting to determined an analytic model of stellar contraction and singularity formation in more than $4D$. The most general form of Vaidya solution for null fluid in Lovelock  theory of gravity have been explored by many authors \cite{13,16,17} and they arrived at the conclusion that the uncovered singularity are feasible for odd dimension for several values of parameters and due to gravitational collapse BH is formed for any value of parameters. Banerjee et al.\cite{14} studied the uncovered singularities in higher dimensional gravitational collapse and concluded that there is a great chance of uncovered singularity. Feinstein \cite{15}  investigated the formation of a black string in a higher dimensional vacuum gravitational collapse. In this paper, the work done by Sharif  and Kausar \cite{12} is extended for $n+2$-dimensional spacetime.
The scheme of the paper is as follow. In section 2, the field
equations are given.  Section 3 is devoted to solutions of the field equations. In section 4, trapped surfaces and apparent horizons are discussed in detail . Finally, the results are summarized in section 5.

\section{LTB Model and Equations of Motion in $f(R)$ gravity}

For the interior region take $n+2$-dimensional non-static spherically symmetric LTB metric is given by
\begin{equation}
 ds^{2}_{-}= dt^{2}-A^{2}dr^{2}-Y^{2}d\Omega^{2},
\label{1}\end{equation}
where the $A=A(r,t)$ and $Y=Y(r,t)$.
\begin{eqnarray}
d\Omega^{2}=\sum^{n}_{k=1}[\prod^{k-1}_{l=1}sin^{2}\theta_{l}]d\theta^{2}_{k}=d\theta_{1}^{2}+sin^{2}\theta_{1}d\theta_{2}^{2}+sin^{2}\theta_{1}sin^{2}\theta_{2}d\theta_{3}^{2}
\nonumber\\
+...+sin^{2}\theta_{1}sin^{2}\theta_{2}sin^{2}\theta_{3}...sin^{2}\theta_{n-1}d\theta_{n}^{2}.\;\;\;\;\;\;\;\;\;\;\;\;\;\;\;\;\;\;\;\;\;\;\;\;
\label{2}\end{eqnarray}
In $f(R)$ gravity equations of motion are \cite{1}-\cite{3}
\begin{equation}
F(R)R_{\pi\chi}-\frac{1}{2}f(R)g_{\pi\chi}-\nabla_{\pi}\nabla{\chi}F(R)+g_{\pi\chi}\nabla^{\sigma}\nabla_{\sigma}F(R)=\kappa T_{\pi\chi}.\;\;\;\;\label{3}
\end{equation}
Here $F(R)=df(R)/dR$, $\nabla_{\pi}$ is the covariant derivative, $T_{\pi\chi}$ is the standard energy momentum tensor and $\kappa$ is the coupling constant. The source perfect fluid is
\begin{equation}
T_{\pi\chi}=(\rho+p)u_{\pi}u_{\chi}-pg_{\pi\chi}, \label{4}
\end{equation}
where $\rho=\rho(r,t)$ is fluid matter density, $p$ is fluid pressure and $u_{\pi}$ is the $n+2$ dimensional velocity vector defined by $u_{\pi}=\delta^{0}_{\pi}$.
For metric (\ref{1}), we get following set of independent partial differential equations are
\begin{eqnarray}
&-&\frac{\ddot{A}}{A}-n\frac{\ddot{Y}}{Y}=\frac{1}{F}[\kappa\rho+\frac{f}{2}]-[-\frac{F^{\prime\prime}}{A^{2}}+\frac{\dot{A}}{A}\dot{F}+\frac{A^{\prime}}{A^{3}}F^{\prime}+n\frac{\dot{Y}}{Y}\dot{F}-n\frac{Y^{\prime}}{A^{2}Y}F^{\prime}],
\label{5}\\
&-&\frac{\ddot{A}}{A}-n\frac{\dot{A}}{A}\frac{\dot{Y}}{Y}+\frac{n}{A^{2}}[\frac{Y^{\prime\prime}}{Y}-\frac{A^{\prime}Y^{\prime}}{AY}]=\frac{1}{F}[\frac{f}{2}-\kappa{p}-A^{2}(\ddot{F}+n\frac{\dot{Y}}{Y}\dot{F}-n\frac{Y^{\prime}}{A^{2}Y}F^{\prime})],
\nonumber\\\label{6}\\
&-&\frac{\ddot{Y}}{Y}-(n-1)(\frac{\dot{Y}}{Y})^{2}-\frac{\dot{A}\dot{Y}}{AY}+\frac{1}{A^{2}}[\frac{Y^{\prime\prime}}{Y}+(n-1)(\frac{Y^{\prime}}{Y})^{2}-\frac{A^{\prime}Y^{\prime}}{AY}-(n-1)(\frac{A}{Y})^{2}]
\nonumber\\
&=&\frac{1}{F}[\frac{f}{2}-\kappa{p}-(\ddot{F}-\frac{F^{\prime\prime}}{A^{2}}+\frac{\dot{A}}{A}\dot{F}+\frac{A^{\prime}}{A^{3}}F^{\prime}+(n-1)\frac{\dot{Y}}{Y}\dot{F}-(n-1)\frac{Y^{\prime}F^{\prime}}{A^{2}Y})],
\label{7}\\
&-&n\frac{\dot{Y^{\prime}}}{Y}+n\frac{\dot{AY^{\prime}}}{AY}=\frac{1}{F}[\dot{F}^{\prime}-\frac{\dot{A}}{A}F^{\prime}].\label{8}
\end{eqnarray}
Here $.$ and $'$ are the partial derivatives with respect to $t$ and $r$, respectively.

Using the scalar perturbation constraints, Cooney et al.\cite{c4} have explored the formation of
 compact objects like Neutron Star in
$f(R)$ gravity. The
Schwarzschild metric cosmological constant has been considered in the external region
that has been matched smoothly with the interior fluid solution using
Darmois junction conditions in a very similar way to GR. According to the authors
\cite{c5,c6} Schwarzschild solution
is the most suitable solution as exterior geometry of the star. In the same way in $f(R)$ gravity
many researchers \cite{f15}-\cite{f181}, \cite{c7,c8} have examined the matching conditions for gravitational collapse.

For exterior region, we consider the  $n+2$-dimensional Schwarzschild metric
\begin{equation}
ds^{2}_{+}=1-\frac{2M}{R}dt^{2}-\frac{1}{1-\frac{2M}{R}}dr^{2}-R^{2}d\Omega^{2}.\label{8}
\end{equation}
The matching conditions require that \\
(1). The first fundamental form of the metrics must be continuous over $\Sigma$ i.e.,
\begin{equation}
(ds^{2}_{+})_{\sum}=(ds^{2}_{-})_{\sum}=(ds^{2})_{\sum},
\label{9}
\end{equation}
2. Also, extrinsic curvature must be continuous over $\Sigma$ i.e.,
\begin{equation}
[K_{cd}]=K^{+}_{cd}-K^{-}_{cd}=0,\;\;\;\;\;\;\;\;\;\;\;\;\;\;\;\;\;\;(i,j=0,2,3...n+1),\label{10}
\end{equation}\\
where $K_{cd}$, the extrinsic curvature tensor is defined as
\begin{equation}
K^{\pm}_{cd}=-n^{\pm}_{\omega}\left( \frac{\partial ^{2}x^{\omega}_{\pm}}{\partial \xi ^{c}\partial \xi ^{d}}%
+{\Gamma _{\gamma\delta}^{\omega}}\frac{{\partial}x^{\gamma}_{\pm}{\partial}x^{\delta}_{\pm}}{{\partial}\xi^{c}{\partial}\xi^{d}}\right),\;\;\;\;\;\;\;\;\; (\omega,\gamma,\delta=0,1,2,3...n+1).\label{11}
\end{equation}
Here
$n^{\pm}_{\omega}$ , $x^{\omega}_{\pm}$ and $\xi^{c}$ denotes the outward, coordinates on $V^{\pm}$, $\Sigma$, respectively.
The equations of hypersurfaces for inner and outer metrics are given by
\begin{eqnarray}
h_{-}(r,t) &=&r-r_{\Sigma }=0  \label{12}, \\
h_{+}(R,T) &=&R-R_{\Sigma }(T)=0,  \label{13}
\end{eqnarray}
where $r_{\Sigma}$ is a constant. Using Eq.(\ref{12})$\ $in Eq.(\ref{1}), the interior metric on the
hypersurface $\Sigma $ takes the following form
\begin{equation}
ds_{-}^{2}=dt^{2}-Y^{2}d\Omega^{2},\label{14}
\end{equation}\\
Also, Eq.(\ref{13})$\ $in Eq. (\ref{8}), we get
\begin{equation}
ds^{2}_{+}=\left(\left(1-\frac{2M}{R}\right)-\frac{1}{\left(1-\frac{2M}{R}\right)}(\frac{dR_{\Sigma}}{dT})^{2}\right)dT^{2}-R^{2}d\Omega^{2}.\label{15}
\end{equation}
For $T$ a timelike coordinate, we assume
\begin{equation}
\left(\left(1-\frac{2M}{R}\right)-\frac{1}{\left(1-\frac{2M}{R}\right)}(\frac{dR_{\Sigma}}{dT})^{2}\right)>0.  \label{16}
\end{equation}\\
Now from junction condition Eq.(\ref{9}), we get
\begin{equation}
R_{\Sigma}=Y(r_{\Sigma},T), \label{17}
\end{equation}
\begin{equation}
[Z(R)-\frac{1}{Z(R)}(\frac{dR_{\Sigma}}{dT})^{2}]^\frac{1}{2}dT=dt,\label{18}
\end{equation}
where $Z(R)=1-\frac{2M}{R}$.
The possible components of $K_{cd}^{\pm}$ are
\begin{eqnarray}
&K_{00}^{-}&=0, \label{21}\\
&K_{22}^{-}&=\csc^{2}\theta_{1}K^{-}_{33}=\csc^{2}\theta_{1}\csc^{2}\theta_{2}K^{-}_{44}=
\nonumber\\
&...=&\csc^{2}\theta_{1}\csc^{2}\theta_{2}...\csc^{2}\theta_{n}K^{-}_{n+1n+1}=(\frac{YY^{\prime}}{A})_{\Sigma},\label{22}\\
&K_{00}^{+}&=[\dot{R}\ddot{T}-\dot{T}\ddot{R}+\frac{3M\dot{R}^{2}\dot{T}}{R(R-2M)}-\frac{\dot{T}^{3}M(R-2M)}{R^{3}}]_{\Sigma},\label{23}\\
&K^{+}_{22}&=\csc^{2}\theta_{1}K^{+}_{33}=\csc^{2}\theta_{1}\csc^{2}\theta_{2}K^{+}_{44}=
\nonumber\\&...=&\csc^{2}\theta_{1}\csc^{2}\theta_{2}...\csc^{2}\theta_{n}K^{+}_{n+1n+1}=(\dot{T}(R-2M))_{\Sigma}.\label{24}
\end{eqnarray}\\
Now from continuity of extrinsic curvature (\ref{10}), it follows that
\begin{equation}
K^{+}_{00}=0,\;\;\;\;\;\;\;\;\;   K_{22}^{+}= K_{22}^{-}.\label{25a}
\end{equation}\\\
Now using Eqs.(\ref{21}-\ref{25a}) along with(\ref{17}) and (\ref{18}), the junction takes the form
\begin{equation}
(A\dot{Y^{\prime}}-\dot{A}Y^{\prime})_{\Sigma}=0,\label{26a}
\end{equation}
\begin{equation}
M=\frac{n-1}{2}Y^{n-1}\left(1-\dot{Y}^{2}-\frac{Y'^2}{A^2}\right)_{\Sigma}.\label{27a}
\end{equation}\\
Eqs.(\ref{17}), (\ref{18}), (\ref{26a}) and (\ref{27a}) are the required conditions for the matching of two regions.

\section{Solution}

We need the explicit value of $A$, for the solution of set of fields equations (\ref{5})-(\ref{8}). It follows from Eq.(\ref{8}) that
\begin{equation}
A=\int\frac{n\dot{Y^{\prime}}F+\dot{F^{\prime}}Y}{nY^{\prime}F+F^{\prime}Y}dt.\label{25}
\end{equation}
To solve above equation, we assume $R=R_{0}$, and $F(R_{0})=canstant$ and this assumption provide $p=p_0$ and $\rho=\rho_0$, here the quantities with subscript $0$ are constant quantities. In modified gravitation theories the
stability of models is tested through the Dolgov-Kawasaki \cite{DK1}
stability criterion
\begin{equation}\label{DK}
F(R)=f_R(R)>0,\quad f_{RR}(R)>0,\quad R{\geq}R_{0}.
\end{equation}
Using above assumptions, the Eqs.(\ref{5})-(\ref{8}) yield
\begin{eqnarray}
&-&\frac{\ddot{A}}{A}-n\frac{\ddot{Y}}{Y}=\frac{1}{F}[\kappa\rho+\frac{f}{2}],\label{25}\\
&-&\frac{\ddot{A}}{A}-n\frac{\dot{A}}{A}\frac{\dot{Y}}{Y}+\frac{n}{A^{2}}[\frac{Y^{\prime\prime}}{Y}-\frac{A^{\prime}Y^{\prime}}{AY}=\frac{1}{F}[\frac{f}{2}-\kappa{p}],\label{26}\\
&-&\frac{\ddot{Y}}{Y}-(n-1)(\frac{\dot{Y}}{Y})^{2}-\frac{\dot{A}\dot{Y}}{AY}+\frac{1}{A^{2}}[\frac{Y^{\prime\prime}}{Y}+(n-1)(\frac{Y^{\prime}}{Y})^{2}-\frac{A^{\prime}Y^{\prime}}{AY}-(n-1)(\frac{A}{Y})^{2}]
\nonumber\\
&=&\frac{1}{F}[\frac{f}{2}-\kappa{p}],\label{27}\\
&-&n\frac{\dot{Y^{\prime}}}{Y}+n\frac{\dot{AY^{\prime}}}{AY}=0.\label{28}
\end{eqnarray}\\
Now from Eq.(\ref{28}), it follows that
\begin{equation}
A(r,t)=\frac{Y^{\prime}(r,t)}{W(r)},\label{29}
\end{equation}\\
where $W=W(r)$. Using Eq.(\ref{29}) in Eqs.(\ref{25})-(\ref{28}), it follows that
\begin{equation}
2\frac{\ddot{Y}}{Y}+(n-1)(\frac{\dot{Y}}{Y})^{2}+(n-1)(\frac{1-W^{2}}{Y^{2}})=-\frac{1}{F(R_{0})}[\frac{8\pi}{n}((n-1)p_{0}-\rho_{0})-\frac{f(R_{0})}{2}].\label{29}
\end{equation}\\
The above equation yields
\begin{equation}
(\dot{Y})^{2}=W^{2}-1+2\frac{m(r)}{Y^{n-1}}-\frac{Y^{2}}{(n+1)F(R_{0})}[\frac{8\pi}{n}((n-1)p_{0}-\rho_{0})-\frac{f(R_{0})}{2}],\label{30}
\end{equation}\\
here $m=m(r)$ and its value is given by
\begin{equation}
m^{\prime}=\frac{8\pi}{nF(R_{0})}((n-1)p_{0}+\rho_{0})Y^{\prime}Y^{n}.\label{31}
\end{equation}
Also, above equation leads to
\begin{equation}
m(r)=\frac{8\pi}{nF(R_{0})}\left(p_{0}(n-1)+\rho_{0}\right)\int (Y'Y^{n} )dr+c(t) ,\label{32}
\end{equation}\\
where $c(t)$ is an arbitrary function of $t$. The function $m(r)$ must be positive. Using second junction condition from Eq.(\ref{27}) and  Eq.(\ref{30}), we get
\begin{equation}
M=(n-1)m(r)-\frac{(n-1)Y^{n+1}}{2(n+1)F(R_{0})}[\frac{8\pi}{n}((n-1)p_{0}-\rho_{0})-\frac{f(R_{0})}{2}].\label{33}
\end{equation}\\
Now using the mass function \cite{9a},  the total energy $M(r,t)$  for the interior spacetime is defined as
\begin{equation}
M(r,t)=\frac{(n-1)Y^{n-1}}{2}[1+g^{\pi\chi}Y_{,\pi}Y_{,\chi}].\label{34}
\end{equation}
Using Eq. (\ref{30}), the mass function becomes
\begin{equation}
M(r,t)=(n-1)m(r)-\frac{(n-1)Y^{n+1}}{2(n+1)F(R_{0})}[\frac{8\pi}{n}((n-1)p_{0}-\rho_{0})-\frac{f(R_{0})}{2}],\label{35}
\end{equation}\\
Now we assume that,
\begin{equation}
\frac{1}{F(R_{0})}[\frac{8\pi}{n}((n-1)p_{0}-\rho_{0})-\frac{f(R_{0})}{2}]>0. \label{34aa}
\end{equation}
and the condition $W(r)=1$ to find the solution, so Eq.(\ref{30}) implies that
\begin{eqnarray}
Y&=&\Bigg(\frac{2(n+1)mF(R_{0})}{\frac{8\pi}{n}((n-1)p_{0}-\rho_{0})-\frac{1}{2}f(R_{0})}\Bigg)^\frac{1}{n+1}\sinh^{\frac{2}{n+1}}\alpha({r,t}),\label{32}\\\nonumber
A&=&\Bigg(\frac{2(n+1)mF(R_{0})}{\frac{8\pi}{n}((n-1)p_{0}-\rho_{0})-\frac{1}{2}f(R_{0})}\Bigg)^\frac{1}{n+1}\Bigg[\frac{m^{\prime}}{(n+1)m}\sinh{\alpha(r,t)}
\\\nonumber
&+&t^{\prime}_{s}(r)\sqrt{\frac{\frac{8\pi}{n}((n-1)p_{0}-\rho_{0})-\frac{1}{2}f(R_{0})}{(n+1)F(R_{0})}}\cosh{\alpha(r,t)}\Bigg]\sinh^{\frac{(1-n)}{(1+n)}}{\alpha(r,t)},\\
\end{eqnarray}\\
where
\begin{equation}
\alpha({r,t})=\sqrt{\frac{(n+1)(\frac{8\pi}{n}((n-1)p_{0}-\rho_{0})-\frac{1}{2}f(R_{0}))}{4F(R_{0})}}[t_{s}(r)-t].\label{33}
\end{equation}\\
We would like to mentioned that in the limit $f(R_{0})\rightarrow \frac{8\pi(p_{0}-\rho_{0})}{n}$, we have Tolman-Bondi solution \cite{19}
\begin{eqnarray}
&Y&=[\frac{(n+1)^{2}m(r)}{2}(t_{s}-t)^{2}]^{\frac{1}{n+1}},\label{34}\\
&A&=\frac{m^{\prime}(t_{s}-t)+2mt^{\prime}_{s}}{[2(n+1)^{n-1}m^{n}(t_{s}-t)^{n-1}]^\frac{1}{n+1}}.\label{35}
\end{eqnarray}
From Eqs.(\ref{DK})and (\ref{34aa}),we get $F(R_0)>0$ and  $f(R_0)<\frac{16\pi}{n}[(n-1)p_0-\rho_0]$. Since Dolgov-Kawasaki \cite{DK1}
stability criterion does not restrict the sign of $f(R_0)$. Therefore for ${\rho_0}>0$, we must have $(n-1)p_0<\rho_0$, it hold for all $n\geq1$ and finally we get $f(R_0) <0$, hence these are the viability conditions for $f(R)$ gravity that must hold through the discussion.
\section{Apparent Horizons}

For spacetime (\ref{1}) the boundary of trapped $n$-sphere is given by
\begin{equation}
g^{\pi\chi}Y,_{\pi}Y,_{\chi}=\frac{A^2\dot{Y}^{2}-Y'}{A^2}=0.\label{34a}
\end{equation}\\
Using Eq.(\ref{30}), above equation yields
\begin{equation}
\frac{1}{F(R_{0})}[\frac{8\pi}{n}((n-1)p_{0}-\rho_{0})-\frac{f(R_{0})}{2})]Y^{n+1}-(n+1)Y^{n-1}+2(n+1)m=0.\label{35}
\end{equation}\\
The values of $Y$ give the boundaries of trapped surfaces which are the apparent horizons. For $f(R_{0})=2(\frac{8\pi}{n}((n-1)p_{0}-\rho_{0})))$, one get $Y=(2m)^{\frac{1}{n-1}}$ which is $n$ Schwarzschild radius. It give de-Sitter horizon, when $m=0$, i.e.,
\begin{equation}
Y=\sqrt{\frac{(n+1)F(R_{0})}{\frac{8\pi}{n}((n-1)p_{0}-\rho_{0})-\frac{f(R_{0})}{2}}}.\label{36}
\end{equation}\\
The approximate solution of Eq.(\ref{35}) up to first order in $m$ and $\frac{1}{F(R_{0})}[\frac{8\pi}{n}((n-1)p_{0}-\rho_{0})-\frac{f(R_{0})}{2})]$, respectively, are given by
\begin{eqnarray}\nonumber
&(Y)_{ch}&=(\frac{(n+1)F(R_{0})}{\frac{8\pi}{n}((n-1)p_{0}-\rho_{0})-\frac{f(R_{0})}{2}})^{\frac{1}{2}}-2((\frac{nF(R_{0})}{8\pi((n-1)p_{0}-\rho_{0})-\frac{1}{2}f(R_{0})})
\nonumber\\\nonumber
&(((n+1)&\frac{8\pi((n-1)p_{0}-\rho_{0})-\frac{1}{2}f(R_{0})}{nF(R_{0})}))^{\frac{n}{2}}+
(\frac{8\pi((n-1)p_{0}-\rho_{0})-\frac{1}{2}f(R_{0})}{n(n-1)(n+1)F(R_{0})})^{\frac{n-2}{2}})m...,\\
\\\nonumber\\
&(Y)_{bh}&=(2m)^{\frac{1}{n-1}}+\frac{1}{F(R_{0})}(\frac{8\pi((n-1)p_{0}-\rho_{0})-\frac{1}{2}f(R_{0})}{n(n-1)(n+1)})(2m)^{\frac{3}{n-1}}f(R_{0})....\label{38}
\end{eqnarray}\\
$(Y)_{ch}$ and $(Y)_{bh}$ are called cosmological and black hole respectively. The existence of $(Y)_{ch}$ is mainly due to the appearance of $f(R)$ term.
Now from Eqs. (\ref{33}) and (\ref{35}), the time for trapped surfaces formation is
\begin{equation}
t_{n}=t_{s}-\sqrt{\frac{4F(R_{0})}{(n+1)[\frac{8\pi}{n}((n-1)p_{0}-\rho_{0})-\frac{f(R_{0})}{2}]}}\sinh^{-1}[\frac{(Y_{n})^{n-1}}{2m(r)}-1]^\frac{1}{2},\;\;\;\;n=1,2.\label{39}
\end{equation}
In the limiting case when $f(R_{0})\rightarrow 2\left(\frac{8\pi}{n}(\rho_{0}-(n-1)p_{0})\right)$, the result coincide to Tolman-Bondi solution
\begin{equation}
t_{n}=t_{s}-\frac{(2^{n}m)^{\frac{1}{n-1}}}{n+1}. \label{40}
\end{equation}
It is clear from Eq.(\ref{39}) formation of trapped surfaces take less time as compared to time of formation of singularity $t=t_{s}$. This implies that horizons form earlier than singularity, hence singularity is covered by the event horizons and end state of gravitational collapse is a BH. The present solutions in $f(R)$ are with the agreement of Oppenheimer-Snyder-Datt models. Hence the end state of gravitational collapse is a BH. From Eq. (\ref{40}) it is to be noted that time for trapped surfaces in higher dimensional Tolman-Bondi spacetime is special case of our present investigation. Further, the $f(R_0)$ term effects the time lag between the formation of trapped surfaces and singularity. The Misner-Sharp mass \cite{9a} has been modified by the $f(R_0)$. Also, the exterior trapped surface, so-called cosmological horizon is due to the presence of $f(R_0)$ term. We explored the physical aspects of the
solutions and found a suitable counterterm in the analytic solutions which avoids
the occurrence of naked singularity during gravitational collapse.

The Dolgov-Kawasaki \cite{DK1}
stability criterion
$F(R)=f_R(R)>0$,~~$R{\geq}R_{0}$ explains that $f(R)$ theory must avoid a ghost state, while the second condition ,$f_{RR}(R)>0,\quad R{\geq}R_{0}$ is introduced to avoid negative mass squared of a scalar-field degree
of freedom. Hence present solutions are ghost free and free of any exotic matter instability caused by the external perturbation \cite{f16}.
This means the final state of gravitational collapse in the present case is not a two phase transition. In other words one may not have any condition to convert a BH into naked singularity. This shows that, in $f(R)$ gravity,
the instability of gravitating system decreases rapidly and system tends to stable state naturally.
This is the important consequence what we expect as $f(R)$ theory modifies the
interaction of gravity by the inclusion of a new scalar
field. We have investigated that $f(R_0)$ plays a dominant role in trapping the collapsing
fields and contributes to the black hole formation. Due to the repulsive nature of scalar force, the $f(R_0)$ term slow down the collapse rate.
\section{Conclusion}
It is particularly interesting to establish the predictions of $f(R)$ theories concerning the gravitational collapse and particularly the collapse time for several astrophysical objects. The outcomes of the analysis of gravitational collapse in $f(R)$ may provide the constraints for the validity of models and be helpful to discard the models which appear against the experimental investigations.
Here, we have examined the gravitational contraction of inhomogeneous perfect fluid in $f(R)$ gravity by considering the metric approach. We have assumed the $n+2$-dimensional spacetime with inhomogeneous and isotropic perfect fluid gravitating source. The $n-$dimensional fluid sphere is taken as interior and Schwarzschild spacetimes as exterior region, respectively.  The general conditions for the smooth matching of two regions have been have been formulated. For the solution of  the field equations,  the assumption of constant curvature is used which implies that pressure and density are constants in this case.
Two physical apparent horizons, black hole horizon and cosmological horizon have been found. We have been shown that trapped surfaces are formed earlier than the singular point of the collapsing sphere, hence singularity is covered by the black hole horizon. This favors the cosmic censorship conjecture.

From equation (\ref{30}), the rate at which fluid sphere collapse occur is given by
\begin{equation}
\ddot{Y}=-\frac{(n-1)m}{Y^{n}}+\frac{Y}{(n+1)F(R_{0})}[\frac{8\pi}{n}((n-1)p_{0}-\rho_{0})-\frac{f(R_{0})}{2}].\label{41}
\end{equation}
For the collapsing process, the acceleration should be negative which is possible when
\begin{equation}
Y<\bigg[-\frac{(n-1)(n+1)mF(R_{0})}{\frac{8\pi}{n}((n-1)p_{0}-\rho_{0})-\frac{f(R_{0})}{2}}\bigg]^{\frac{1}{n+1}}.\label{42}
\end{equation}
It is evident from Eq. (\ref{41}) that the $f(R_{0})$ slows down the collapsing process (as mentioned in \cite {f1}) when $\frac{1}{F(R_{0})}(\frac{8\pi}{n}((n-1)p_{0}-\rho_{0})-\frac{f(R_{0})}{2})<{0}$ is satisfied. Further, due to $f(R_{0})$ there exists two physical horizon namely block hole horizon and cosmological horizon. We would like to point out that for $n=2$, our results match to the results of Sharif and Kausar \cite{12}.

As mentioned earlier (in the introduction) that there are two types of solutions concerning the gravitational collapse in $f(R)$  gravity, one predicts naked singularity and other BH. In \cite{f5, f6, f10}, the final state of gravitational collapse in $f(R)$ gravity is a naked singularity, while in \cite{f8, f9, f19,  f11, f12} BH has been found as a final outcome of collapse in $f(R)$ gravity. Thus our results favors the investigations of \cite{f8, f9, f19, f11, f12} and may considered as one example of Oppenheimer-Snyder-Datt models in $f(R)$ gravity. The $f(R_0)$ term slows down the process of gravitational collapse this favors the finding of Ref.\cite {f1}. Finally, we would like to mentioned that the results of this paper can be extended in the frame work of other modified theories of gravity, like $f(T)$, $f(G)$, $f(R,G)$ and $f(R,T)$ etc.

\section*{Acknowledgement}

The constructive comments and suggestions of anonymous referee are highly acknowledged.

\end{document}